\documentclass[12pt]{article}
\usepackage{amssymb}
\usepackage{epsfig}

\catcode`\@=11
\def\numberbysection{\@addtoreset{equation}{section}
        \def\theequation{\thesection.\arabic{equation}}}

\def\beq{\begin{equation}}
\def\eeq{\end{equation}}
\numberbysection
\begin{document}
\begin{titlepage}
\begin{center}
\hfill \\
\vskip 1.in {\Large \bf Translation in momentum space and minimal length} \vskip 0.5in P. Valtancoli
\\[.2in]
{\em Dipartimento di Fisica, Polo Scientifico Universit\'a di Firenze \\
and INFN, Sezione di Firenze (Italy)\\
Via G. Sansone 1, 50019 Sesto Fiorentino, Italy}
\end{center}
\vskip .5in
\begin{abstract}
We show that in presence of the Snyder algebra the notion of translation in momentum space is modified to a formula similar to the relativistic addition of velocities. These results confirm the strict connection between Snyder algebra and the Lorentz group.
\end{abstract}
\medskip
\end{titlepage}
\pagenumbering{arabic}
\section{Introduction}

Among the possible methods to quantize gravity we can mention the introduction of a minimal length in physical theories. This idea was introduced in 1947 by Harland Snyder \cite{1} but was later abandoned due both to the difficulty of introducing a minimal length in quantum field theory and to the success of the renormalization theory for the standard model.

The fact that gravity is the only non-renormalizable physical theory has however left open the possibility of defining a physical theory in the presence of a minimal length, a natural cutoff for the ultraviolet divergences that afflict quantum field theory.

In general, the simplest non-commutative field theories are defined by introducing a star-product between the fields (a type of non-commutative product). However, reconciling Snyder's algebra with quantum field theory remains a subject of considerable difficulty. In this work we try to give a new direction with which to face this long-standing problem.

 Normally the Fourier transform is used to establish a relationship between the functions defined on a Minkowski space and the operators defined on a Hilbert space.
In the case of Snyder's algebra the operator $ e^{ i k_\alpha \hat{x}^\alpha } $ ( which is central in defining the mapping ) can be considered as a deformation of the translation operator in momentum space. However, as we will calculate later, this deformation introduces some fictitious singularities and leads to poorly defined results.

 To overcome this difficulty, we introduce an alternative deformation of the translation operator in momentum space which is free of singularities. We anticipate that a close relationship is obtained between this deformation and the formula for adding the velocities in special relativity.This confirms the close connection between
 Snyder's algebra and the Lorentz group, while the approach with the Fourier transform seems to be incompatible with the structure of Snyder' s algebra.

\section{Noncommutative field theories and the star product}

Non-commutative field theories have been the subject of recent studies \cite{2}-\cite{3}-\cite{4}. In the case that space-time is non-commutative in the sense that

\beq [ \ \hat{x}^\mu, \hat{x}^\nu \ ] \ = \ i \ \theta^{\mu\nu} \label{21}
\eeq

where $ \theta^{\mu\nu} $ is a constant matrix, there is a correspondence between functions $ f $ defined on the Minkowski space and operators $ F $ defined on a Hilbert space given by :

\beq F( \hat{x} ) \ = \ \int \ \frac{d^4 x}{(2\pi)^2} \ e^{ i k \cdot \hat{x} } \ \tilde{f} ( k ) \label{22}
\eeq

where $ \tilde{f}$ is linked to a function $f$ in the position space:

\beq \tilde{f} ( k ) \ = \ \int \ \frac{d^4 x}{(2\pi)^2} \ e^{ -i k \cdot x } \ f ( x ) \label{23}\eeq

 In addition, the following inversion formula applies:

 \beq \tilde{f} ( k ) \ = \ Tr ( e^{ - i k \cdot \hat{x} } \ F ( \hat{x} ) ) \label{24} \eeq

 where the trace is defined by:

 \beq Tr (A) \ = \ \lim_{\Lambda \rightarrow \infty} \ \frac{(2\pi)^2}{\Lambda^4} \ \int^{\Lambda} \ d^4 q \ < q | A | q > \label{25}\eeq

 where  the $ | q > $ are momentum eigenvectors. This inversion formula works because the following identity holds:

 \beq < h | \ e^{ i q \cdot \hat{x} } \ e^{ - i k \cdot \hat{x} } \ | h > \ = \ \delta ( q-k ) \label{26}\eeq

 Similarly, the product of operators maps to the star-product of functions

 \beq F \ G \ \leftrightarrow f \star g \label{27}\eeq

 where the star-product is defined by

 \beq f \star g \ = \ \lim_{x' \rightarrow x} \ e^{\frac{i}{2} \ \theta^{\alpha\beta} \ \partial_\alpha \partial^{'}_\beta } \ f(x) \ g(x')
  \label{28} \eeq

  However, our interest is in Snyder's algebra, defined in terms of the compact variable $ \rho^i $ as follows:

  \beq \hat{x}^i \ = \ i \hbar \sqrt{ \ 1 \ - \ \beta \ \rho^2 \ } \ \frac{\partial}{\partial \rho^i} \ \ \ \ \ \ \ p^i \ = \ \frac{\rho^i}{\sqrt{ \ 1 \ - \ \beta \ \rho^2 \ }} \ \ \ \
  \ \ \ 0 < \rho^2 < \frac{1}{\beta} \label{29} \eeq

  In this case it is still possible to define a relationship between functions $ f $ and operators $ F $ through (\ref{22}) and (\ref{23}) ( $ f \rightarrow F $ ) but it is not possible to easily reverse this relationship ( $ F \rightarrow f $ ) ( see for details \cite{2} ). This fact makes it difficult to calculate the star product in the presence of Snyder's algebra.

  \section{Translation in momentum space}

  Let's analyze the following operator in detail

  \beq e^{ i k^\alpha \cdot \hat{x}_\alpha } \ = \ e^{ - \ k^\alpha \ \sqrt{ \ 1 \ - \ \beta\rho^2 \ } \frac{\partial}{\partial \rho_\alpha} } \label{31} \eeq

In the limit $ \beta \rightarrow 0 $ this operator is nothing else than the translation operator in momentum space :

\beq | \rho_0 + k > \ = \ e^{ - \ k^\alpha \ \frac{\partial}{\partial \rho_\alpha} } \ | \rho_0 > \label{32} \eeq

If $ \beta \neq 0 $ we can exactly calculate its action on the operator $ \rho^\alpha $:

\beq  e^{- i \ k \cdot \hat{x} } \ \rho^\alpha \ e^{ i \ k \cdot \hat{x} } \ = \ \rho^\alpha \ + \ k^\alpha \ \left[ \sqrt{1-\beta\rho^2} \ \frac{ \sin ( \sqrt{\beta k^2} ) }{ \sqrt{\beta k^2} } \ + \ \beta \ ( k \cdot \rho ) \ \left( \ \frac{ \cos ( \sqrt{\beta k^2} ) \ - \ 1 }{\beta k^2} \ \right) \
\right] \label{33} \eeq

from which we get

\beq | {\rho'}^\alpha_0 > \ = \ e^{ i \ k \cdot \hat{x} } \ | \rho^\alpha_0 > \ = \ | \ \rho^\alpha_0 \ + \ k^\alpha \ \left[ \sqrt{1-\beta\rho^2_0} \ \frac{ \sin ( \sqrt{\beta k^2} ) }{ \sqrt{\beta k^2} } \ + \ \beta \ ( k \cdot \rho_0 ) \ \left( \ \frac{ \cos ( \sqrt{\beta k^2} ) \ - \ 1 }{\beta k^2} \ \right) \ \right] > \label{34} \eeq

The problem we raise now is that the possible values of $ \rho_0^\alpha $ must meet the condition

\beq 0 < \rho^2_0 < \frac{1}{\beta} \label{35} \eeq

while the transformed $ {\rho'}_0^{\alpha} $ does not meet this requirement. Hence the operator (\ref{31}) takes out of the allowed space and is poorly defined. For example in $ d=1 $ a finite translation ( $ \sqrt{\beta \ k^2} \ = \ \alpha $ ) can bring a finite momentum ( $ \beta \rho^2_0 \ = \ cos^2 \alpha $ ) to an infinite momentum ( $ \beta \rho'^2_0 \ = \ 1 $ ), a rather unphysical behaviour. Furthermore the product of two operators of the type (\ref{31}) $ e^{ i \ h \cdot \hat{x} } \ e^{ i \ k \cdot \hat{x} } $ is very complicated.

In general one can define a generic deformation for the translation in momentum space using the following formula:

\beq ( \ x' , \ \rho' \ ) \ = \ e^{ \ k^\alpha \ f( \beta \rho^2 ) \ \frac{\partial}{\partial \rho^\alpha } } \ ( \ x, \ \rho \ ) \  e^{ - \ k^\alpha \ f( \beta \rho^2 ) \ \frac{\partial}{\partial \rho^\alpha } } \label{35} \eeq

but for a generic $ f( \beta \rho^2 ) $ it is not warranted that the transformed $ {\rho'}^\alpha $ satisfies the constraint $ 0 < {\rho'}^2 < \frac{1}{\beta} $, leading to singularities in the transformed momentum $ p' $ for some finite value of $ \rho $ and  $ k $.

In the next chapter we will define a new deformation which doesn't introduce fictitious singularities in the mapping $ ( \ x, \ \rho \ ) \ \rightarrow \ ( \ x', \ \rho' \ ) $ and which reduces to a translation in the momentum space in the limit $ \beta \rightarrow 0 $.

\section{One dimensional case}

 We require that the mapping $ T : \rho \rightarrow \rho' $ meets the following two requirements:

 i) if $ \rho $ belongs to the range $ 0 < \rho^2 < \frac{1}{\beta} $ then also the transformed $ \rho'$ does the same

\beq 0 < {\rho'}^2 < \frac{1}{\beta} \label{41} \eeq

ii) in the limit $ \beta \rightarrow 0 $ the mapping $ T $ reduces to simple translation $ \rho' = \rho + k $.

These two requirements are met by the following mapping

\beq \rho \rightarrow \rho' \ = \ \frac{\rho \ + \ k}{ 1 \ + \ \beta \ k \ \rho } \label{42} \eeq

In particular it is true that

\beq ( \ 1 \ - \ \beta \ \rho^2 \ ) \ \rightarrow ( \ 1 \ - \ \beta \ {\rho'}^2 \ ) \ = \ \frac{( \ 1 \ - \ \beta \ k^2 \ )( \ 1 \ -\ \beta \ \rho^2 \ )}{( \ 1 \ + \ \beta \ k \ \rho \ )^2} \label{43} \eeq

so $ 0 < {\rho'}^2 < \frac{1}{\beta} $ is valid if $ 0 < k^2 < \frac{1}{\beta} $ and $ 0 < \rho^2 < \frac{1}{\beta} $. We obtain as a consequence that the translation parameter $ k $ is also limited in the same interval.

To obtain a symmetry of the algebra

\beq [ \ x , \rho \ ] \ = \ i \hbar \ \sqrt{ \ 1 \ - \ \beta \ \rho^2 \ } \label{44} \eeq

we also have to change $ x  \rightarrow x' $:

\beq x \rightarrow x' \ = \ \frac{ 1 \ + \ \beta \ k \ \rho}{\sqrt{ 1 \ - \ \beta \ k^2}} \ x \label{45} \eeq

Thus it is ensured that

\beq [ \ x , \rho \ ] \ = \ i \hbar \ \sqrt{ \ 1 \ - \ \beta \ \rho^2 \ } \ \ \rightarrow  \ \ [ \ x' , \rho' \ ] \ = \ i \hbar \ \sqrt{ \ 1 \ - \ \beta \ {\rho'}^2 \ } \label{46} \eeq

Going from the reduced variable $ \rho $ to the momentum variable $ p $ we obtain:

\begin{eqnarray}
p \rightarrow p' & = & \frac{1}{ \sqrt{ 1 \ - \ \beta \ k^2 }} \ [ \ p \ + \ k \ \sqrt{ 1 \ + \ \beta \ p^2 } \ ] \nonumber \\
x \rightarrow x' & = & \frac{1}{ \sqrt{ 1 \ - \ \beta \ k^2 }}  \left[ \ 1 \ + \frac{\beta \ k \ p }{ \sqrt{ 1 \ + \ \beta \ p^2} } \ \right] \ x \label{47}
\end{eqnarray}

The composition of two transformations of this type is simple:

\beq \rho \rightarrow \rho' \ = \ \frac{\rho \ + \ k}{ 1 \ + \ \beta \ k \ \rho} \rightarrow \rho'' \ = \ \frac{\rho' \ + \ h}{1 \ + \ \beta \ h \ \rho'} \ = \
\frac{\rho \ + \ \widetilde{h}}{1 \ + \ \beta \ \widetilde{h} \ \rho}
\label{48} \eeq

where

\beq \widetilde{h} \ = \ \frac{ h \ + \ k}{1 \ + \ \beta \ h \ k} \label{49} \eeq

Obviously also $ \widetilde{h}^2 < \frac{1}{\beta} $ since $ h^2 < \frac{1}{\beta} $ and $ k^2 < \frac{1}{\beta} $.

\section{Generalization to the Snyder algebra}

The generalization to Snyder's algebra is trivial if one remembers how the addition of velocities is done in special relativity;

\beq {\rho'}^\alpha \ = \ \frac{1}{( \ 1 \ + \ \beta \ \rho \cdot k \ ) } \left\{ \ \rho^\alpha \ \left[ \ 1 \ + \ \left( \ 1 \ - \ \sqrt{1-\beta \rho^2} \ \right) \frac{\rho \cdot k}{\rho^2} \ \right] \ + \ \sqrt{1-\beta \rho^2} \ k^\alpha \ \right\} \label{51} \eeq

The following properties are valid:

\begin{eqnarray}
\rho' \cdot \rho & = & \frac{\rho^2 \ + \ \rho \cdot k}{( 1 \ + \ \beta \ \rho \cdot k )} \nonumber \\
\sqrt{ 1 \ - \ \beta \ \rho^2 } & \rightarrow & \sqrt{ 1 \ - \ \beta \ {\rho'}^2 } \ = \ \frac{\sqrt{ 1 \ - \ \beta \ k^2 } }{( 1 \ + \ \beta \ \rho \cdot k ) } \
\sqrt{ 1 \ - \ \beta \ \rho^2} \label{52} \end{eqnarray}

To obtain a symmetry of Snyder's algebra

\beq [ \ x^\alpha, \rho_\beta \ ] \ = \ i \hbar \ \sqrt{ 1 \ - \ \beta \ \rho^2} \ \delta_{\alpha\beta} \ \ \ \rightarrow \ \ \
[ \ {x'}^\alpha, \rho'_\beta \ ] \ = \ i \hbar \ \sqrt{ 1 \ - \ \beta \ {\rho'}^2} \ \delta_{\alpha\beta} \label{53} \eeq

we must transform $ x^\alpha \rightarrow {x'}^\alpha $ as follows :

\beq {x'}^\alpha \ = \ i \hbar \ \sqrt{ 1 \ - \ \beta \ {\rho'}^2} \frac{\partial}{\partial \rho'_\alpha} \ = \ \frac{\sqrt{ 1 \ - \ \beta \ k^2 } }{( 1 \ + \ \beta \ \rho \cdot k ) }
\ \sum_\beta \frac{\partial \rho_\beta}{\partial \rho'_\alpha} \ x^\beta \label{54} \eeq

We note that this definition allows a simple composition of these transformations:

\beq {x''}^\alpha \ = \ i \hbar \ \sqrt{ 1 \ - \ \beta \ {\rho''}^2} \frac{\partial}{\partial \rho''_\alpha} \ = \ \frac{\sqrt{ 1 \ - \ \beta \ {\rho''}^2}}{\sqrt{ 1 \ - \ \beta \ {\rho'}^2}}
\ \sum_\beta \frac{\partial \rho'_\beta}{\partial \rho''_\alpha} \ {x'}^\beta \ = \ \frac{\sqrt{ 1 \ - \ \beta \ {\rho''}^2}}{\sqrt{ 1 \ - \ \beta \ {\rho}^2}}
\ \sum_\beta \frac{\partial \rho_\beta}{\partial \rho''_\alpha} \ x^\beta \label{55}
\eeq

The calculation of this matrix of partial derivatives

\beq f_{\alpha\beta} \ = \ \frac{\partial\rho_\beta}{\partial \rho'_\alpha} \label{56} \eeq

is complicated. Let's first calculate:

\beq \frac{\partial \rho'_\beta}{\partial\rho_\alpha} \ = \ \eta_{\alpha\beta} \ A_1 + \rho_\alpha \rho_\beta \ A_2 + k_\alpha k_\beta \ A_3 + \rho_\alpha k_\beta \ A_4 \ + \
k_\alpha \rho_\beta \ A_5 \label{57} \eeq

where

\begin{eqnarray}
A_1 & = & \frac{1}{( \ 1 \ + \ \beta \ \rho \cdot k \ )} \ \left[ \ 1 \ + \ \left( \ 1 \ - \ \sqrt{1-\beta \rho^2} \ \right) \frac{\rho \cdot k}{\rho^2} \ \right] \nonumber \\
A_2 & = & \frac{1}{( \ 1 \ + \ \beta \ \rho \cdot k \ )} \ \left[ \frac{\beta \ \rho \ \cdot \ k}{ \sqrt{ 1 \ - \ \beta \ \rho^2 } \ \rho^2} \ - \ 2 \left( \ 1 \ - \ \sqrt{1-\beta \rho^2} \ \right) \ \frac{\rho \cdot k}{\rho^4} \right] \nonumber \\
A_3 & = & - \frac{\beta \ \sqrt{1-\beta \rho^2} }{( \ 1 \ + \ \beta \ \rho \cdot k \ )^2} \nonumber \\
A_4 & = & - \frac{\beta}{( \ 1 \ + \ \beta \ \rho \cdot k \ ) \ \sqrt{1-\beta \rho^2} } \nonumber \\
A_5 & = & - \frac{\beta}{( \ 1 \ + \ \beta \ \rho \cdot k \ )^2} \ + \ \frac{\left( \ 1 \ - \ \sqrt{1-\beta \rho^2} \ \right)}{\rho^2 \ ( \ 1 \ + \ \beta \ \rho \cdot k \ )^2 }
\label{58} \end{eqnarray}

Let us define

\beq f_{\alpha\beta} \ = \ \eta_{\alpha\beta} \ B_1 + \rho_\alpha \rho_\beta \ B_2 + k_\alpha k_\beta \ B_3 + \rho_\alpha k_\beta \ B_4 \ + \
k_\alpha \rho_\beta \ B_5 \label{59} \eeq

The coefficients $ B_i $ can be obtained from $ A_i $. Finally we get the following expressions:

\begin{eqnarray}
B_1 & = & \frac{1}{A_1} \nonumber \\
B_2 & = & - \frac{A_2}{\Delta} - \frac{k^2}{A_1 \Delta} \ ( \ A_2 A_3 \ - \ A_4 A_5 \ ) \nonumber \\
B_3 & = & - \frac{A_3}{\Delta} - \frac{\rho^2}{A_1 \Delta} \ ( \ A_2 A_3 \ - \ A_4 A_5 \ ) \nonumber \\
B_4 & = & - \frac{A_4}{\Delta} + \frac{\rho \cdot k}{A_1 \Delta} \ ( \ A_2 A_3 \ - \ A_4 A_5 \ ) \nonumber \\
B_5 & = & - \frac{A_5}{\Delta} + \frac{\rho \cdot k}{A_1 \Delta} \ ( \ A_2 A_3 \ - \ A_4 A_5 \ )
\label{510} \end{eqnarray}

where the denominator is

\begin{eqnarray} \Delta & = & ( \ A_1 \ + \ \rho^2 \ A_2 \ + \ ( \rho \cdot k ) \ A_5 \ ) ( \ A_1 \ + \ k^2 \ A_3 \ + \ ( \rho \cdot k ) \ A_4 \ ) \ - \nonumber \\
& - & ( \ k^2 \ A_5 \ + \ ( \rho \cdot k ) \ A_2 \ ) ( \ \rho^2 \ A_4 \ + \ ( \rho \cdot k ) \ A_3 ) \label{511} \end{eqnarray}

\section{Connection with $\beta$-canonical transformations}

It is possible to show that the transformation (\ref{54}) also satisfies the condition of $\beta$-canonical transformation ( see \cite{5} ), therefore it is well defined. In the $ 1d$ case we simply have to prove that

\beq \left( \ \frac{\partial x'}{\partial x} \ \frac{\partial \rho'}{\partial \rho} \ - \ \frac{\partial \rho'}{\partial x} \ \frac{\partial x'}{\partial \rho} \ \right) \ \sqrt{ 1 \ - \ \beta \ \rho^2 \ } \ = \ \sqrt{ 1 \ - \ \beta \ {\rho'}^2 \ } \label{61} \eeq

where

\beq \rho' \ = \ \frac{\rho \ + \ k}{ 1 \ + \ \beta \ k \ \rho } \ \ \ \ x' \ = \ \frac{1 \ + \ \beta \ k \ \rho}{\sqrt{1 \ - \ \beta \ k^2}} \ x \label{62} \eeq

Verification reduces to the following identity

\beq \frac{\partial \rho'}{\partial \rho} \ = \ \frac{1 \ - \ \beta \ k^2}{( \ 1 \ + \ \beta \ k \ \rho \ )^2} \label{63} \eeq

which is true.

In general we have to show that:

\beq \left\{ \ x'_i, \rho'_j \ \right\}_{\{ x_i, \rho_j \}} \ = \ \sqrt{1 \ - \ \beta \ {\rho'}^2} \ \delta_{ij} \label{64} \eeq

\beq \left\{ \ x'_i, x'_j \ \right\}_{\{ x_i, \rho_j \}} \ = \ \beta \left( \ \frac{x'_i \ \rho'_j \ - \ x'_j \ \rho'_i}{\sqrt{ 1 \ -  \ \beta \ {\rho'}^2}} \ \right) \label{65} \eeq

where the bracket is modified:

\begin{eqnarray} \{ \ u_i , v_j \ \}_{\{ q_i, w_j \}} & = & \sqrt{ 1 \ - \ \beta \ w^2 } \ \sum^{n}_{k=1} \ \left( \ \frac{\partial u_i}{\partial q_k} \ \frac{\partial v_j}{\partial w_k}
 \ - \  \frac{\partial u_j}{\partial q_k} \ \frac{\partial v_i}{\partial w_k} \ \right) \nonumber \\
 & + & \beta \ \sum^n_{l,m \ =  \ 1} \ \left( \ \frac{ q_l \ w_m \ - \ q_m \ w_l }{\sqrt{1 \ - \ \beta \ w^2}} \ \right) \ \frac{\partial u_i}{\partial q_l} \ \frac{\partial v_j}{\partial q_m} \label{66} \end{eqnarray}

and

\beq x'_i \ = \ \frac{ \sqrt{ 1 \ - \ \beta \ {\rho'}^2} }{ \sqrt{ 1 \ - \ \beta \ \rho^2} } \  \sum_k \ \frac{\partial \rho_k}{\partial \rho'_i} \ x_k \label{67} \eeq

We first prove equation (\ref{64}). We obtain as an intermediate step

\beq \sum^n_{k=1} \ \left( \ \frac{\partial x'_i}{\partial x_k} \ \frac{\partial \rho'_j}{\partial \rho_k} \ \right) \ = \ \frac{ \sqrt{ 1 \ - \ \beta \ {\rho'}^2} }{ \sqrt{ 1 \ - \ \beta \ \rho^2} } \ \delta_{ij} \label{68} \eeq

which is true.

Let us prove equation (\ref{65}). We can rewrite the $\beta$-canonical bracket as

\begin{eqnarray}
\{ x'_i , x_j \} & = & \alpha_1 \ + \ \alpha_2 \nonumber \\
\alpha_1 & = & \ \sqrt{ 1 \ - \ \beta \ \rho^2 } \ \sum^{n}_{k=1} \ \left( \ \frac{\partial x'_i}{\partial x_k} \ \frac{\partial x'_j}{\partial \rho_k}
 \ - \  \frac{\partial x'_j}{\partial \rho_k} \ \frac{\partial x'_i}{\partial x_k} \ \right) \ \nonumber \\
 \alpha_2 & = & \beta \ \sum^n_{l,m \ =  \ 1} \ \left( \ \frac{ x_l \ \rho_m \ - \ x_m \ \rho_l }{\sqrt{1 \ - \ \beta \ \rho^2}} \ \right) \ \frac{\partial x'_i}{\partial x_l} \ \frac{\partial x'_j}{\partial x_m} \label{69}\end{eqnarray}

Calculating the derivatives is easy. We obtain

\begin{eqnarray} \frac{\partial x'_i}{\partial x_k} & = & \frac{\sqrt{ 1 \ - \ \beta \ k^2 }}{( \ 1 \ + \ \beta \ \rho \ \cdot \ k \ )} \ \frac{\partial \rho_k}{\partial \rho'_i}
\nonumber \\ \frac{\partial x'_i}{\partial \rho_k} & = &  \ - \ \frac{\beta\  k_k \ x'_i }{( \ 1 \ + \ \beta \ \rho \ \cdot \ k \ ) } \ + \
\frac{\sqrt{ 1 \ - \ \beta \ k^2 }}{( \ 1 \ + \ \beta \ \rho \ \cdot \ k \ )} \ \sum^n_{l=1} \ \frac{\partial}{\partial \rho_k} \ \left( \ \frac{\partial \rho_l}{\partial \rho'_i}
\ \right) \ x_l \label{610} \end{eqnarray}

The first term $ \alpha_1 $ gives rise to

\beq \alpha_1 \ = \ \frac{\sqrt{ 1 \ - \ \beta \ {\rho'}^2 }}{( \ 1 \ + \ \beta \ \rho \cdot k \ )} \  \ \left[ \ - x'_j \ \frac{\partial ( \beta \rho \cdot k ) }{ \partial \rho'_i } \ - \ ( i \leftrightarrow j ) \ \right]
\label{611} \eeq

The second term proportional to $ \beta $ gives rise to

\beq \alpha_2 \ = \ \frac{\beta \ ( \ 1 \ - \ \beta \ k^2 \ )}{ 2 \ ( \ 1 \ + \ \beta \ \rho \ \cdot \ k \ )^2 \ \sqrt{ 1 \ - \ \beta \ \rho^2 }} \ \left[ \ x_l \ \frac{\partial \rho_l}{\partial \rho'_i} \ \frac{\partial \rho^2}{\partial \rho'_j} \ - \ ( i \leftrightarrow j ) \ \right] \label{612} \eeq

Using this identity

\beq ( \ 1 \ - \ \beta \ \rho^2 \ ) \ = \ \frac{( \ 1 \ + \ \beta \ \rho \ \cdot k \ )^2 }{( \ 1 \ - \ \beta \ k^2\ )} \ ( \ 1 \ - \ \beta \ {\rho'}^2 \ )\label{613} \eeq

we obtain

\beq \alpha_2 \ = \ \beta \left( \ \frac{x'_i \ \rho'_j \ - \ x'_j \ \rho'_i}{\sqrt{ 1 \ -  \ \beta \ {\rho'}^2}} \right) \ + \ \frac{\sqrt{ 1 \ - \ \beta \ {\rho'}^2 }}{( \ 1 \ + \ \beta \ \rho \cdot k \ )} \  \ \left[ \ - x'_i \ \frac{\partial ( \beta \rho \cdot k ) }{ \partial \rho'_j } \ - \ ( i \leftrightarrow j ) \right] \label{614} \eeq

Adding the two contributions equation (\ref{65}) is verified.

\section{A solvable example}

To define a mapping between functions and operators in the case of the deformation described in this paper, we limit ourselves to the soluble case in 1d. We must first find the explicit representation:

 \begin{eqnarray}
  \rho & \rightarrow & \rho' \ = \ \frac{ \rho \ + \ k }{ 1 \ + \ \beta \ k \ \rho \ } \ = \ e^H \ \rho \ e^{-H} \nonumber \\
  \frac{\partial}{\partial \rho} & \rightarrow  & \frac{\partial}{\partial \rho'} \ = \ e^H \ \frac{\partial}{\partial \rho}\ e^{-H}
 \label{71} \end{eqnarray}

 where $ H \ = \ f(\rho) \frac{\partial}{\partial \rho} $ is a linear operator.

The $ H $ operator must satisfy the condition:

\beq f(\rho') \ \frac{\partial}{\partial \rho'} \ = \ f(\rho) \ \frac{\partial}{\partial \rho} \label{72} \eeq

from which we obtain

\beq f( \rho' ) \ = \ \frac{( \ 1 \ - \ \beta \ k^2 \ )}{( \ 1 \ + \ \beta \ k \ \rho \ )^2} \ f( \rho ) \label{73} \eeq

 The general solution of this equation is

 \beq f( \rho ) \ = \ c(k) \ ( \ 1 \ - \ \beta \ \rho^2 \ ) \label{74} \eeq

 where $ c $ is a constant dependent on $ k $.

 The constant $ c(k) $ can be obtained with a perturbative calculation

 \beq c(k) \ = \ k \ ( \ 1 \ + \ \frac{1}{3} \ \beta \ k^2 \ + \ \frac{1}{5} \ \beta^2 \ k^4 \ + ... \ ) \ = \ \frac{1}{\sqrt{\beta}} \ \tanh^{-1}{(\sqrt{\beta k^2})} \label{75} \eeq

 or from the condition that the composition of two transformations

 \beq e^{H(h)} \ e^{H(k)} \ = \ e^{H(\widetilde{h})} \label{76} \eeq

 where  $ \widetilde{h} $ is defined by the equation ( \ref{49} ).

 Let us notice that  $c(k) = \eta(k)$ is the rapidity function of special relativity:

 \beq \eta(k) \ + \ \eta(h) \ = \ \eta(\widetilde{h}) \label{77} \eeq

  At this point we notice the substantial difference between our solution

  \beq H \ = \ \eta(k) \ ( \ 1 \ - \ \beta \ \rho^2 \ ) \ \frac{\partial}{\partial\rho} \label{78} \eeq

  and the operator (\ref{31}).

  Furthermore, the rapidity variable $ \eta $ can extend to infinity while the variable $ k $ is bounded $ 0 < k^2 < \frac{1}{\beta} $.

 Also the bounded variable $ \rho $ can be replaced with the variable

 \beq \rho \ \rightarrow \ y \ = \ \frac{1}{\sqrt{\beta}} \ \tanh^{-1} ( \sqrt{\beta\rho^2} ) \label{79} \eeq

 which is unbounded. Then the translation operator defined in this work takes the standard form

\beq e^{i \ \eta \ \hat{x}} \label{710} \eeq

where the rapidity $ \eta $ takes the role of momentum variable in the Fourier transform and $ \hat{x} \ = \ - i \frac{\partial}{\partial y} $.

So we can try to define a mapping between functions and operators of the form:

\begin{eqnarray}
f(x) & = & \int \ d\eta \ e^{i \ \eta \ x} \ \widetilde{f} ( \eta ) \nonumber \\
F( \hat{x} ) & = & \int \ d\eta \ e^{i \ \eta \ \hat{x}}  \ \widetilde{f} ( \eta )
\label{711} \end{eqnarray}

which is the basis for defining a field theory in noncommutative geometry.

 In the general case there is certainly an operator $ H $ such that

 \beq \rho^a \ \rightarrow \ \rho'^a \ = \ e^H \ \rho^a \  e^{-H} \label{712} \eeq

 but it is difficult to find a closed form for the linear operator $ H $ as we did in case 1d. We leave this discussion for future work.

\section{Conclusions}

Normally non-commutative field theories are defined in terms of a star-product that modifies fields in interaction. In the case of Snyder algebra we have criticized this method based on the Fourier transform, because applying it we obtain a deformation of the translation operator in momentum space that introduces a fictitious singularity for some finite value of $ k $ and $ \rho $.

In this work we have introduced a correct deformation of the translation operator in momentum space constructed starting from the formula of the addition of velocities in special relativity.

 This deformation is a true symmetry of Snyder's algebra. Confirmation of this is the verification that this deformation has the property of being a $\beta$-canonical transformation (introduced as symmetry of the path-integral in our previous work \cite {5}).

Therefore we find a close connection between Snyder algebra and the Lorentz group, while the Fourier transform method is more suitable for non-commutative theories in which the translation in momentum space is the standard one.

We hope that the knowledge of this deformation can stimulate the construction of a new method to define quantum field theory in the presence of Snyder's algebra.


\begin{thebibliography}{999}
\bibitem{1} Snyder H. S., Phys. Rev. {\bf 71} (1947) 38.
\bibitem{2} Licht A. L., arXiv:hep-th/0512134.
\bibitem{3} Battisti M. V. and Meljanac S., Phys. Rev. D {\bf 82} (2010) 024028, arXiv:1003.2108.
\bibitem{4} Girelli F. and Livine E. R., JHEP {\bf 1103} (2011) 132, arXiv:1004.0621
\bibitem{5} Valtancoli P., J. Math. Phys. {\bf 56} (2015) 12 122107, arXiv:1510.01651.

\end{thebibliography}
\end{document}